\documentstyle[preprint,aps,graphics]{revtex}
\begin{document}


\title{Fluctuations of Complex Networks:
\\Electrical Properties of Single Protein Nanodevices} 

\author{C. Pennetta $^a$, V. Akimov$^a$, E. Alfinito$^a$,
L. Reggiani$^a$, and G. Gomila$^b$}
%

\address{
$^a$INFM, National Nanotechnology Laboratory, 
Via Arnesano, 73100, Italy\\ 
and Dipartimento di Ingegneria dell'Innovazione, \\
Universit\`a di Lecce, Via Arnesano, Lecce 73100, Italy.\\
$^b$ Department d'Electronica and Research Centre for Bioelectronics 
and \\Nanobioscience, Universitat de Barcelona, C/ Josep Samitier 1-5, 08028\\
Barcelona, Spain.}


\date{\today}
\maketitle
\begin{abstract}
\noindent
We present for the first time a complex network approach to the study of the 
electrical properties of single protein devices. In particular, we consider an
electronic nanobiosensor based on a G-protein coupled receptor. By adopting a 
coarse grain description, the protein is modeled as a complex network of 
elementary impedances. The positions of the alpha-carbon atoms of each amino 
acid are taken as the nodes of the network. The amino acids are assumed to 
interact electrically among them. Consequently, a link is drawn between any 
pair of nodes neighboring in space within a given distance and an elementary 
impedance is associated with each link. The value of this impedance 
can be related to the physical and chemical properties of the amino acid pair 
and to their relative distance. Accordingly, the conformational changes of the
receptor induced by the capture of the ligand, are translated into a variation
of its electrical properties. Stochastic fluctuations in the value of the 
elementary impedances of the network, which mimic different  physical effects,
have also been considered. Preliminary results concerning the impedance 
spectrum of the network and its fluctuations are presented and discussed for 
different values of the model parameters.
\end{abstract}

\keywords{Complex networks, G protein coupled receptors, 
Impedance fluctuations, Nanobiodevices}

\maketitle



\thanks{Corresponding authors e-mail: cecilia.pennetta@unile.it }
\date{\today}
\maketitle 
%

\section{INTRODUCTION}
\label{sect:intro}  
Originated as an evolution of the Erd\"os and R\'eny theory of random graphs, 
the study of complex networks has started in '98-'99 with the seminal works of 
Watts and Strogatz \cite{strogatz98} and Barabasi and Albert \cite{barabasi99}.Thenceforth, it has fastly grown, becoming one of the most lively research 
fields \cite{albert,strogatz,jeong,callaway,goltsev,alon,argollo,barabasi04}. 
The reasons of this boom rely on the fact that the concepts and techniques 
which are introduced and developed in the theory of complex networks offer 
very powerful tools for the study of many real-life systems like: 
communication networks, social behaviors, metabolic and cellular networks, 
electric power grids, etc 
\cite{albert,strogatz,jeong,callaway,goltsev,alon,argollo,barabasi04}. 

On the other hand, the attention to develop new hybrid molecular electronic 
devices and, in particular, of ultra miniaturized single-molecule devices, has
also grown tumultuously, since the first single-molecule experiment performed 
30 years ago on biomolecules \cite{hladky}, in view of a huge amount of 
technological applications \cite{alon,wu,joachim,bayley,firestein}. Therefore,
nowadays a very large number of experimental and computational works are 
devoted to this mandatory issue \cite{alon,barabasi04,wu,joachim,bayley,firestein,xie,bezrukov,lameh,shacham,vaidehi,elrod}. However, in many cases,
the atomic complexity of bioelectronic systems makes arduous the task of 
performing a microscopic study of their electrical properties.

Here we present for the first time a complex network approach to the study of 
the electrical properties of a two terminal device that is a promising 
candidate for the realization of single protein nanobiosensors \cite{spot}. 
The device under consideration consists of a G protein-coupled receptor (GPCR) 
\cite{lameh,gether,shacham,vaidehi,elrod} embedded in a lipid bilayer 
and contacted to two ideal electrodes. In particular, we will show results for 
bovine rhodopsin \cite{joachim,palcz,yeagle,choi,sakmar} (a photonic receptor)
although the results can be easily extended to other receptors belonging to 
the GPCR family, as for instance olfactory receptors (ORs)
\cite{wu,bayley,firestein,pilpel,crasto,liu}. In fact, all GPCRs share the 
seven helices trans-membrane structure, shown schematically in Fig. 1. 
The seven trans-membrane helices (TH), are interconnected by extracellular 
(EC) and intracellular (IC) loops. Additionally, there are an extracellular 
terminal chain (N terminus) and an intracellular terminal chain  (C terminus) 
\cite{lameh,gether,shacham,vaidehi,elrod,palcz,yeagle}.
We note that the understanding of the electrical properties of GPCRs is 
crucial to the purpose of developing single-protein electronic biosensors.
However, the information about their structure at atomic level is generally 
missing. By contrast, this information is available for rhodopsin 
\cite{palcz,yeagle,vaidehi,pdb}. Indeed, in this case it is possible to pack 
together the molecules in a crystal structure and to determine the atomic 
coordinates by diffraction experiments \cite{palcz,yeagle}. 

Therefore, by taking advantage of the common topology of the peptidic chain of
rhodopsin and other GPCRs, we have developed the following "coarse grain" 
approach to the electrical properties of these kind of receptors. The receptor
is modeled as a complex network of elementary impedances. The positions of the
alpha-carbon atoms, $C_{\alpha}$, of each amino acid (residue) are taken as 
the reference positions of the network nodes. The amino acids are assumed to 
interact electrically among them. Charge transfers between neighboring 
residues \cite{xie,yang} and/or changes of their electronic polarization 
\cite{song} are taken to affect the state of these interactions. Therefore, a 
link is drawn between any pair of nodes neighboring in space within a given a 
distance and an elementary impedance is associated with each link. This 
elementary impedance is taken as the impedance of a parallel RC circuit. Its 
value can be related to the physical and chemical properties of the amino acid
pair \cite{song} and to their relative distance. The conformational changes of
the receptor induced by the photons or by the capture of a ligand
\cite{firestein,kobilka2,kobilka,yang}, are then translated into a variation 
of its electrical properties. Stochastic fluctuations in the value of the 
elementary impedances of the network, which mimic different physical effects, 
can be also acconted for. The impedance spectrum of the network and its 
fluctuations are studied and analyzed for different values of the parameters 
introduced in the model.

\section{MODEL AND RESULTS}
\label{sect:model}  
In the following we will focus on the modelization of the electrical 
properties of bovine rhodopsin, the best known GPCR  
\cite{lameh,gether,shacham,vaidehi,elrod,palcz,yeagle,choi,sakmar}. 
To this purpose, we consider a two terminal device made by a single molecule 
of rhodopsin embedded in a small portion of membrane, placed within a suitable
environment (physiological buffer solution) and inserted between two metallic 
contacts, to which an AC voltage can be applied. The device is sketched in 
Fig. 2, where, without loss of generality, a vertical configuration has been 
assumed for the electrodes (other configurations can also be considered). The 
left and right hand side of Fig. 2, refer to the basic state and to the most 
stable light-activated state of rhodopsin, respectively. Rhodopsin in the last 
state is referred to as metarhodopsin II \cite{choi}. The atomic coordinates 
of bovine rhodopsin have been taken from Protein Data Bank (PDB) \cite{pdb} 
where data of several independent experiments are present as standard PDB 
files (we have used PDB IDs: 1F88 \cite{palcz} and 1JFP \cite{yeagle}). There 
is also an engineered data for metarhodopsin II (PDB ID 1LN6)\cite{choi}.

The device is then modeled as a complex network of elementary impedances  that 
is constructed according to the following procedure. First, from the PDB file 
we extract the spatial coordinates of the $C_{\alpha}$ atom of each amino acid
(348 for bovine rhodopsin). The positions of these atoms are taken to coincide
with the nodes of the network. We assume that the amino acids interact 
electrically among them through charge transfer processes between neighboring 
residues \cite{xie,yang} and/or changes of their electronic polarization 
\cite{song}. Accordingly, as illustrated in Fig. 3(a), a link is drawn between
any pair of nodes which are neighboring in space within a given distance 
$d \equiv 2R_a$, where $R_a$ represents an interaction radius. Moreover, we 
introduce two extra nodes that are associated with the electrodes 
(contact-nodes), as shown in Fig. 3(b). These contact-nodes are linked to a 
given set of amino acids, depending on the particular geometry of the contacts
in the real device (each electrode is linked at least to one amino acid). 
The environment is not taken into account at this preliminary stage of the 
model. Then, an impedance is assigned to each link. We take this elementary 
impedance as the impedance of a RC parallel circuit (the most usual 
equivalent passive AC circuit). Then, we denote as $Z_{i,j}$, $R_{i,j}$, 
$C_{i,j}$, respectively the impedance, the resistance and the capacitance 
associated with the link between the $i$-th and $j$-th nodes (see Fig. 4). 

Different expressions can be adopted for the determination of $Z_{i,j}$. Here,
we have considered three possibilities corresponding to an increasing level of
complexity. The first possibility, model (i), is the simplest one: all the 
impedances are taken to be equal: $Z_{i,j}=Z_0$. The second possibility, model
(ii), consists in assuming that $R_{i,j}$ is a simple ohmic resistor and 
$C_{i,j}$ a planar homogeneous capacitor, therefore: $R_{i,j} \propto l_{i,j}$
and $C_{i,j} \propto 1/l_{i,j}$, where $l_{i,j}$ is the distance between the
nodes $i$ and $j$. Consequently, $Z_{i,j}$ takes the expression:
%
\begin{equation}
Z_{i,j}= \Bigl({ l_{i,j}\over A } \Bigr)
{1\over (\rho^{-1} + i \epsilon\epsilon_0\omega)}
\label{eq:zij}
\end{equation} 
%
where $A$ is the cross-sectional area of the capacitor, $\rho$ the resistivity
of the resistor, $\epsilon$ the relative dielectric constant, $\epsilon_0$ the 
dielectric constant of vacuum and $\omega$ the frequency of the external AC 
voltage. The third choice, model (iii), consists in taking the cross-sectional 
area of the resistor and of the capacitor equal to the area of the 
cross-section defined by the overlap of the two spheres in Fig. 4. 
In this case, $Z_{i,j}$ becomes:
%
\begin{equation}
Z_{i,j}={l_{i,j}\over (R_a^2 -l_{i,j}^2/4)} 
{1\over (\rho^{-1} + i \epsilon\epsilon_0\omega)}
\label{eq:zijsmoth}
\end{equation} 
%
Of course, an improvement in the description of the receptor is expected if
the physical and chemical properties of the different amino acids are 
accounted for in the expression of $Z_{i,j}$. At a simplest level, this can be
done by taking $\epsilon$ and/or $\rho$ dependent on the indices $i$ and $j$. 
To this purpose, here, we have assumed the following expression for the 
dielectric constant of the capacitor associated with the pair of amino acid 
$i$ and $j$: $\epsilon_{i,j}=1+g[(\alpha_i+\alpha_j)/2 - 1]$, where $\alpha_i$
and $\alpha_j$ are the intrinsic polarizabilities of the corresponding amino 
acids, given in Ref. \cite{song} The factor $g=4.647$ in the previous 
expression of $\epsilon_{i,j}$ has been introduced to the purpose of
obtaining values of $\epsilon_{i,j}$ distributed between 1 and 80 (vacuum and 
water) proportionally to $(\alpha_i+\alpha_j)/2$.

A comparison of the three models previously introduced has been carried out
by taking $\epsilon_{i,j}$ independent of the indices $i$ and $j$ and it 
concerns with the dependence of the network impedance, $Z$, 
on the interaction radius. The network impedance is calculated by solving 
Kirchhoff's node  equations, as in Ref. \cite{rammal} We note that in the 
present case of an irregular network with complex topology, the use of node 
equations is particularly convenient with respect to the use of loop 
equations.  
Figure 5 shows the modulus of the network impedance $|Z|$, calculated by using
models (i), (ii) and (iii), as a function of the interaction radius.
The systematic decrease of $|Z|$ at increasing values of $R_a$ reflects the 
increasing importance of parallel with respect to series connections. 
One can see that the curves of the first two models show a step-like behavior 
related to the sharp discontinuity in the value of $|Z|$  when $R_a$  becomes 
equal to $l_{i,j}$. On the contrary, by removing the impedance value 
discontinuity, the curve obtained by using model (iii) shows a continuous 
behavior. Furthermore, model (iii) appears to be more sensitive to a variation
of the number of links in the networks. Therefore, in the following we will 
discuss results obtained by using model (iii). 

Figure 6 shows the degree distribution, i.e. the distribution function of the 
node connectivities, for different values of the interaction radius, ranging
from $2 \div 6$ \AA. The degree distribution is found to be Poissonian-like 
for all values of $R_a$. This is a signature of a random network 
\cite{jeong,albert}. We note that the width of the distribution becomes 
progressively larger at increasing values of $R_a$. This trend will saturate 
at a certain value $R_{a,max}$ of the interaction radius, which corresponds 
to a network where each nodes is connected with all the 
other nodes. At saturation, for values of  $R_a >R_{a,max}$, the degree of a 
node (number of neighbors of a node) becomes equals to $(N_a -1)/2$, where 
$N_a$ is the number of amino acids, independently of $R_a$.
  
Figure 7 reports the total number of links in the network, $N$, as a function 
of the interaction radius, for the basic state of rhodopsin (black dashed 
curve) and for the activated state metarhodopsin II (gray dotted curve). 
We can see that for an arbitrary value of $R_a$, $N$ is different in the two 
states. In fact, even if the primary structure of the protein remains the 
same, the conformational change induced by the photon modifies the distances 
between $C_{\alpha}$ atoms. As a consequence, for a given value of the 
interaction radius, the network changes: in the activated configuration of the
receptor some new links will arise and some other will disappear. Furthermore,
Fig. 7 displays the relative increment of the number of links, 
$\Delta N/\Delta R_a$, as a function of $R_a$: the solid curves, black and 
gray, show this quantity respectively in the basic and in the activated states
of rhodopsin. Overall, from the behavior of $N$ and of $\Delta N/\Delta R_a$, 
we conclude that the sensitivity of the network to conformational changes is 
maximum when $R_a \approx 10-18$ \AA. It must be noted that this dependence of
the total network impedance on conformational changes is further emphasized by
the fact that the elementary impedances $Z_{i,j}$ are taken to be dependent on
the distances $l_{i,j}$.

%
  

Impedance spectroscopy measurements are frequently used to investigate the 
electrical properties of self-assembled layers of biomolecules lying on 
functionalized metallic supports \cite{joachim}. Even in the case under 
consideration, a single-molecule device, we expect useful information from 
this kind of technique. Therefore, we have calculated the impedance of the 
network as a function of the frequency of the applied AC voltage. Figure 8 
shows the Nyquist plot of the impedance corresponding to the rhodopsin 
network. Precisely, the figure displays the minus imaginary part of the 
network impedance versus the real part, where both these quantities are 
calculated in the frequency range $0 \div 1$ KHz. The two curves reported in 
the figure are obtained by taking $R_a=2$ \AA \, (dashed curve) and 
$R_a=12.5$ \AA \, (solid curve). In both cases it is $\rho=10^9$ $\Omega$m 
while the amplitude of the applied voltage, $V_0$, is $V_0=1$ V. Moreover, the
real and the imaginary part of the impedance have been normalized to the 
static value of the real part, $Re[Z(\omega=0)]$, which takes the values 
$Re[Z(\omega=0)]=302$ G$\Omega$ and $Re[Z(\omega=0)]=1.67$ M$\Omega$, 
respectively for $R_a=2$ \AA \, and $R_a=12.5$ \AA. 
As a general trend, when $R_a \geq 5$ \AA, the shape of the Nyquist plot is 
indistinguishable from that corresponding to a single RC parallel circuit 
(semi-circle). By contrast, when $R_a \le 5$ \AA \, the Nyquist plot deviates 
from this behavior. In particular, when $R_a=2$ \AA, the degree of most of 
the nodes of the network is $2$ (see Fig. 6) and the series combination of 
elementary impedances $Z_{i,j}$ becomes predominant in the network structure. 
In other terms, changes in the shape of the Nyquist plot are only detected in 
the sequential limit. Of course, the value of $Re[Z(\omega=0)]$ depends
on both $R_a$ and $\rho$. Therefore, impedance spectroscopy measurements
will allow the identification of the values of the parameters to be used in 
the modelization of the receptor. Furthermore, the Nyquist plot is a 
convenient tool to detect the presence of series resistances in the contacts
and other spurious phenomena. 

Figure 9 displays a comparison between the Nyquist plot corresponding to the 
rhodopsin network and that corresponding to  the metarhodopsin II network.
Precisely, the solid curve is obtained for rhodopsin while the dot-dashed 
curve for metarhodopsin II. In both cases we have taken $R_a=12.5$ \AA, 
$\rho=10^9$ $\Omega$m, and $V_0=1$ V. The real and the imaginary part of the 
impedance are normalized to the static value of the real part of the network 
impedance in the metarhodopsin state: $Re[Z(\omega=0)]=2.04$ M$\Omega$.
The figure shows that the conformational change of the receptor due to the 
the photon implies a significant variation in the Nyquist curve, in principle 
detectable by impedance spectroscopy measurements. This result, is of 
particular relevance in view of the application of the model to ORs. 

As a next step we have considered the possibility of introducing stochastic 
fluctuations in the interaction network. In fact, fluctuations of the number 
of links or fluctuations of the elementary impedances $Z_{i,j}$, must be 
allowed to account for the fluctuations of the electrical properties of the 
receptor. These last can arise from different mechanisms, like thermal 
fluctuations of the atomic positions, fluctuations in the charge transport 
and/or in the polarization state of the amino acids, etc. 
As a starting level, we have extended the approach already developed for the 
case of regular-lattice networks \cite{prl85} to the case of the rhodopsin 
interaction network. Therefore, starting from the rhodopsin network previously
constructed, that can be called perfect network, we have assumed the existence 
of two random processes, consisting in the breaking and in the recovery of the
links. The impedance of a broken link is taken greater than that an active link
for a factor of $10^8$. By defining $W_b$ and $W_r$ as the probabilities of 
occurrence of the two processes, the network either reaches a
steady state or breaks, depending on the values of $W_b$ and $W_r$. In the 
first case, the fraction of broken links, $p$, and the network impedance, $Z$,
fluctuate around their respective average values $<p>$ and $<Z>$. In the 
second case, a path of broken links spans the network. This second possibility
occurs when $p$ reaches a characteristic value, percolation threshold, that
depends on the network structure \cite{stauffer}. 

By Monte Carlo simulation we have calculated the evolution of the rhodopsin 
interaction network when the two random  processes described above are present.
Figure 10 reports the results obtained for different values of $W_b$ and 
$W_r$ and which correspond to steady states of the network. Precisely, Fig. 10
shows the modulus of the network impedance versus time (measured in units of
iteration steps). In this case we have used the following values of the 
parameters: $R_a=5$ \AA, $\rho=10^6$ $\Omega$m,  $V_0=1$ V, while the 
frequency is $\omega=8$ Hz. The short dashed curve, the solid gray and the 
solid black curves are obtained by taking:  
$W_b=2 \times 10^{-3}, 5 \times 10^{-3}, 6 \times 10^{-3}$, respectively. 
The different curves evidence that at increasing value of $W_b$, the 
fluctuations become more and more relevant, denoting an increasing instability
of the network. 

\section{CONCLUSIONS}
\label{sect:concl}
We have presented a complex network approach to the study of the electrical 
properties of single protein devices. In particular, we have considered a two 
terminal device consisting of a G protein coupled receptor (rhodopsin) 
embedded in a lipid bilayer and contacted to two ideal electrodes. A coarse 
grain description has been developed for the description of the electrical 
properties of the receptor, which is modeled as a network of elementary 
impedances. The conformational changes of the receptor induced by the capture 
of the ligand (photon), are then translated into a variation of the network 
impedance. The role played by the different parameters of the model on the 
network structure and on the impedance spectral properties has been studied. 
Furthermore, stochastic fluctuations in the value of the elementary impedances
of the network, which mimic different physical effects, have been considered. 

\acknowledgments     
This work has been performed within the SPOT-NOSED project IST-2001-38899 of 
EC. Moreover, partial support from the cofin-03 project ``Modelli e misure di
rumore in nanostrutture'' financed by Italian MIUR is also gratefully 
acknowledged. We thank Edith Pajot, Roland Salesse and their group at INRA 
(Jouy en Josas, France) for providing us with valuable information about the 
properties of GPCRs. Moreover, we thank Nicole Jaffrezic (IFOS, Lyon, France), 
Marco Sampietro (Politecnico, Milano, Italy), J. Baussells (CNM, Barcelona, 
Spain) and J. Samitier (UB-CBEN, Barcelona, Spain) for helpful discussions.


 \vfill\break\noindent

FIGURE CAPTIONS

\vspace{0.5truecm}\noindent
Fig. 1 . Schematic representation of a G protein coupled receptor in its 
   environment. The gray snake-like curve represents the seven helices 
   trans-membrane receptor, which is embedded in the cellular membrane 
   (lipid bilayer), shown as the black double lines, which separates the 
   extracellular region from the intracellular region (cytoplasm). The ligand 
   captured by the receptor is shown as the gray triangle. The N and 
   C terminus represent respectively the extracellular and the intracellular 
   terminal chains

\vspace{0.5truecm}\noindent
Fig. 2 . Receptor contacted between two electrodes in a vertical 
   configuration. On the left, schematic representation of the basic state of 
   the rhodopsin, on the rigth of the activated state. The meaning of the 
   letter is the following: 1, N-terminus; 2, trans-membrane core; 3, 
   C-terminus. 

\vspace{0.5truecm}\noindent
Fig. 3 .a) Interaction network associated with a hypothetical protein
   made of 11 residues: the circles show the nodes positioned at the alpha
   carbon atom of each amino acid, the lines represent the links arising from 
   electrical interactions; b) addition of two extra nodes associated
   with the electrodes.

\vspace{0.5truecm}\noindent
Fig. 4 . A link is drawn between two nodes, associated with the amino acids 
   i and j, when the distance between their respective alpha carbon atoms
   is less than twice the interaction radius. An elementary 
   impedance taken as the impedance of a parallel RC circuit, 
   is attributed to each link.

\vspace{0.5truecm}\noindent
Fig. 5 . Modulus of the network impedance as a function of the 
   interaction radius: the dashed-dotted curve is obtained from model 
   (i), the solid curve from model (ii) and the long-dashed curve from model 
   (iii). The impedance is expressed in arbitrary units, the interaction 
   radius is in \AA.

\vspace{0.5truecm}\noindent
Fig. 6 .  Degree distribution (distribution function of the node 
   connectivities) for different values of the interaction radius.  
   Precisely, the radius is: 2, 3, 4, 5, 6 \AA.

\vspace{0.5truecm}\noindent
Fig. 7 . Total number of links in the network, N, and its relative 
   increment as a function of the interaction radius. The black dashed 
   curve shows N in the basic state of rhodopsin and the gray dotted curve 
   in the activated state (metarhodopsin II). The solid curves, black and 
   gray, shows the relative increment of N respectively
   in the basic state and in the activated states of rhodopsin. The 
   interaction radius is expressed in \AA.

\vspace{0.5truecm}\noindent
Fig. 8 . Nyquist plot of the network impedance:
   the dashed curve corresponds to a value of the interaction radius equals to
   2. \AA \, and the solid one to a value of 12.5 \AA. The real and the 
   imaginary part of the impedance are normalized to the static value of the 
   real part, which takes the values 302 G Ohm and 1.67 M Ohm, respectively 
   when the interaction radius is 2 and 12.5 \AA. Both curves correspond to 
   the basic state of rhodopsin and are obtained by taking a resistivity 
   of 1 G Ohm m.

\vspace{0.5truecm}\noindent
Fig. 9 . Nyquist plot of the network impedance:
   the solid curve is obtained for rhodopsin by taking the interaction radius 
   equals to 12.5 \AA \, and a resistivity of 1 G Ohm m, the dot-dashed curve 
   is obtained for metarhodopsin by taking the same values of the parameters. 
   The real and the imaginary part of the impedance are normalized to the 
   static value of the real part of the network impedance in the metarhodopsin
   state which takes the value 2.04 M Ohm.

\vspace{0.5truecm}\noindent
Fig. 10 . Fluctuations of the modulus of the impedance versus times for 
   different values of the breaking and recovery probabilities,
   as specified in the figure. The impedance is expressed in Ohm, the 
   time in simulation steps.

\clearpage
\resizebox{5.5in}{!}{\includegraphics{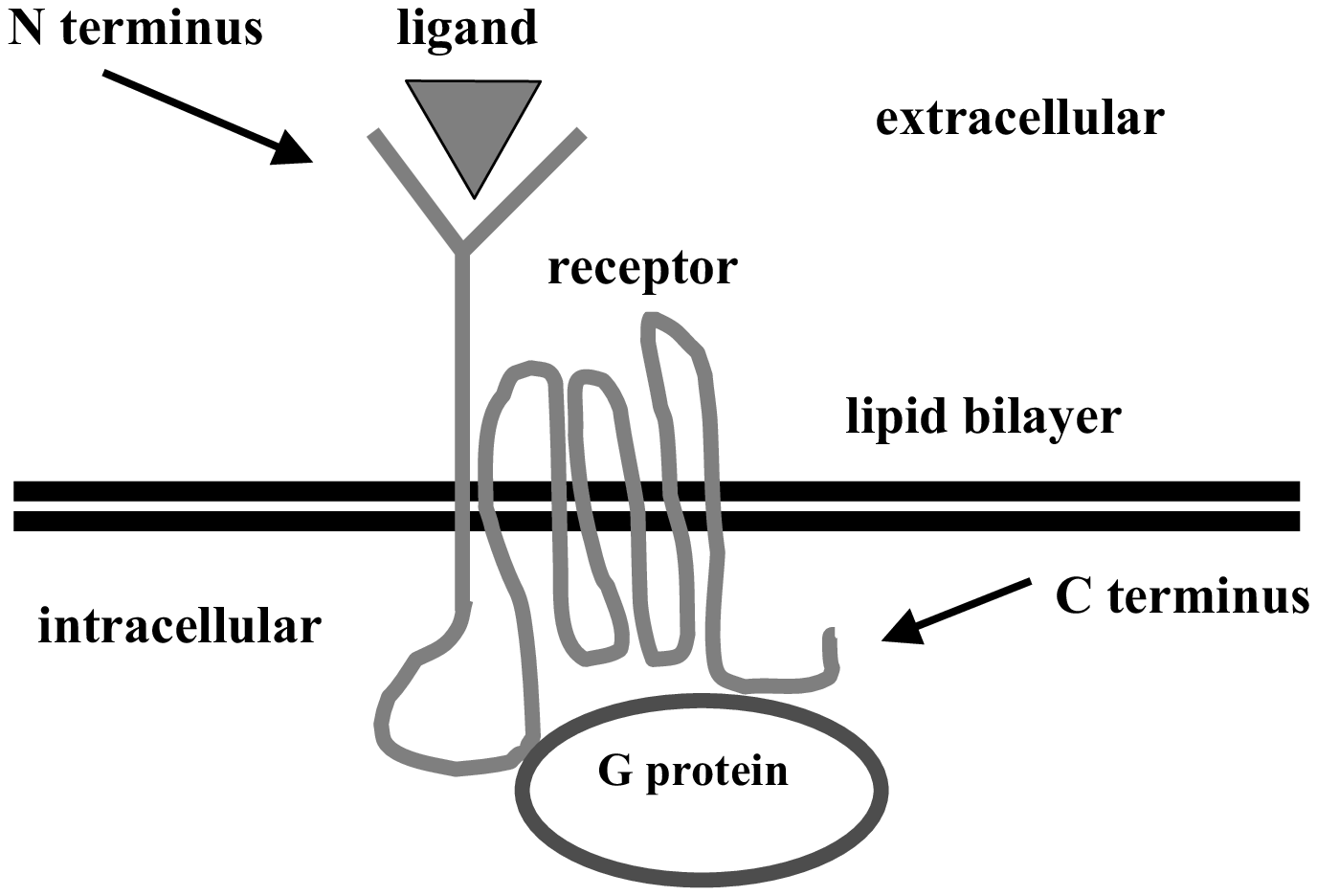}}
\clearpage
\resizebox{5.5in}{!}{\includegraphics{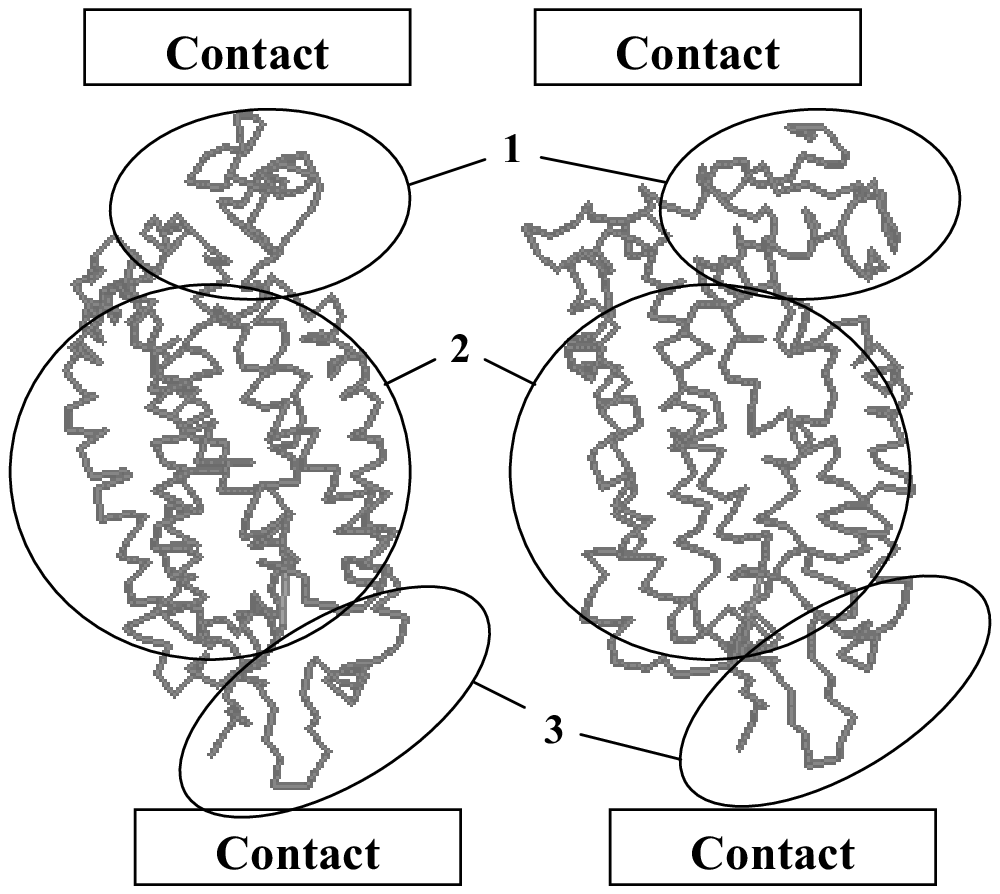}}
\clearpage
\resizebox{5.5in}{!}{\includegraphics{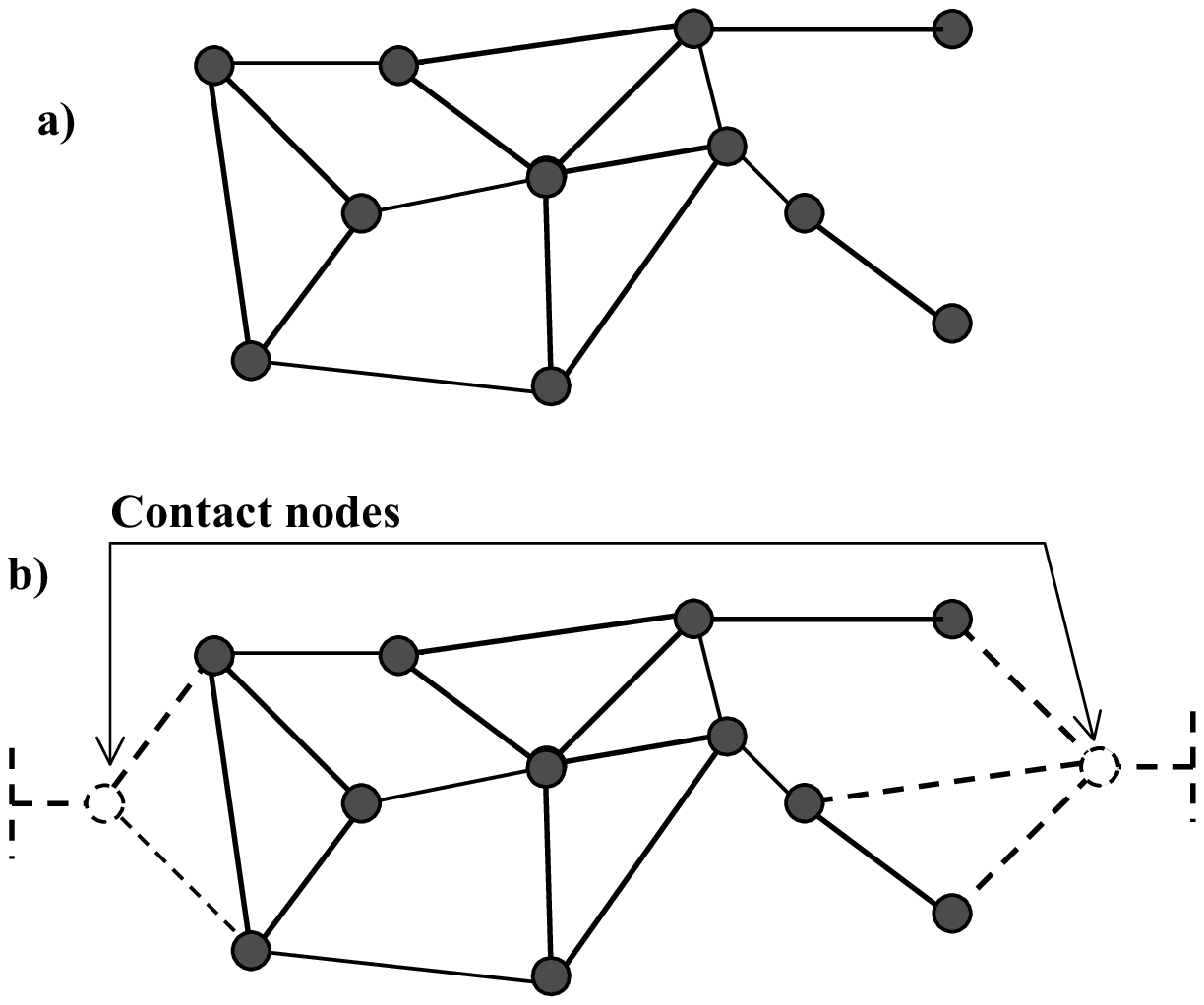}}
\clearpage
\resizebox{5.5in}{!}{\includegraphics{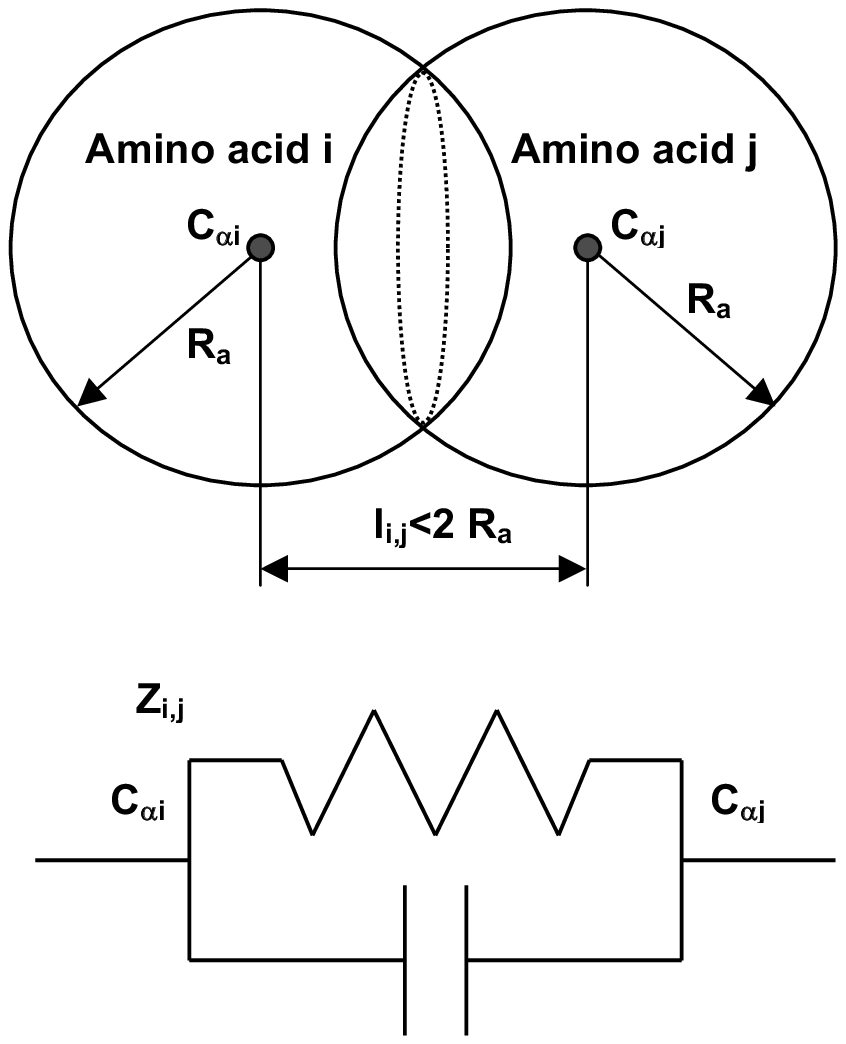}}
\clearpage
\resizebox{5.5in}{!}{\includegraphics{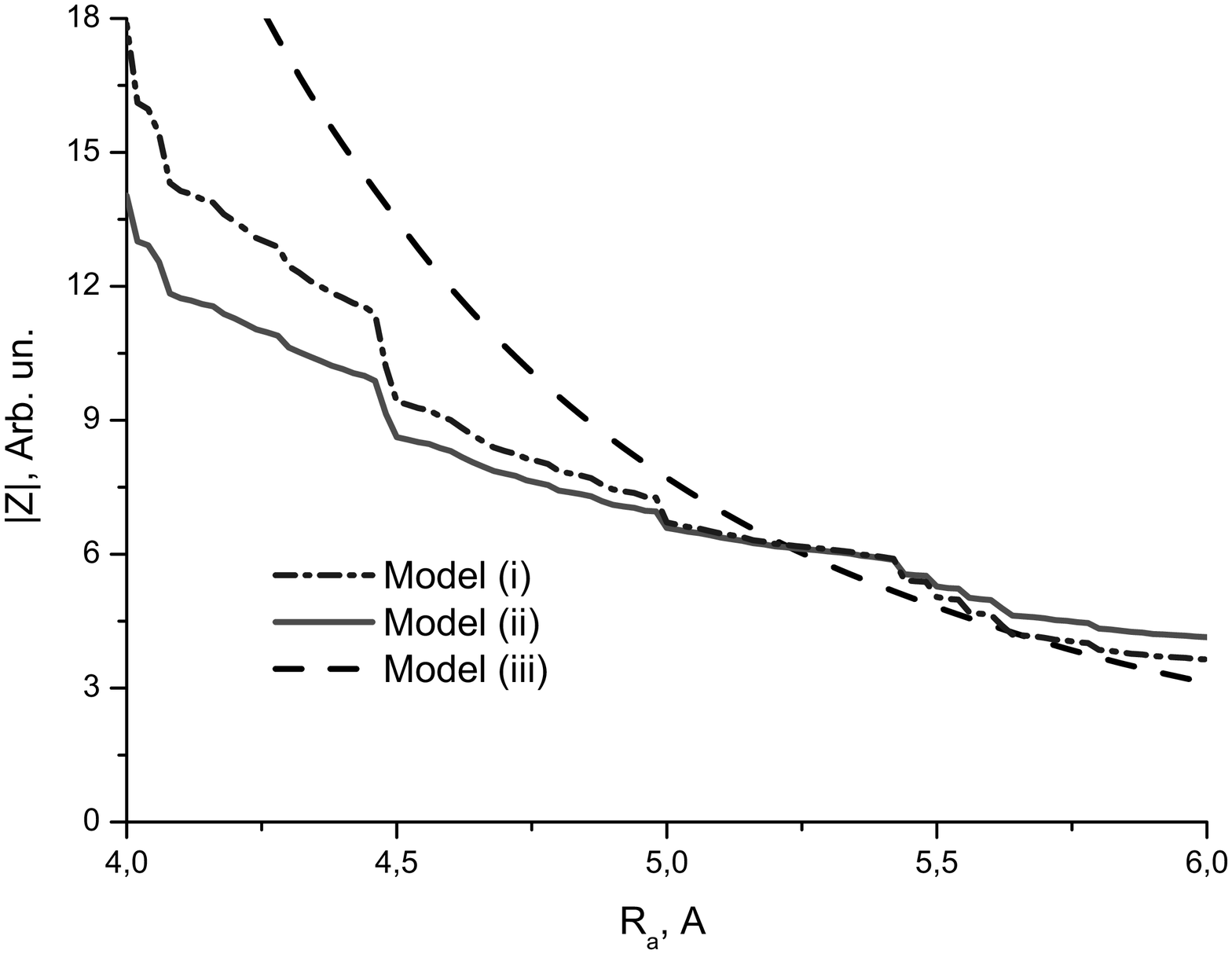}}
\clearpage
\resizebox{5.5in}{!}{\includegraphics{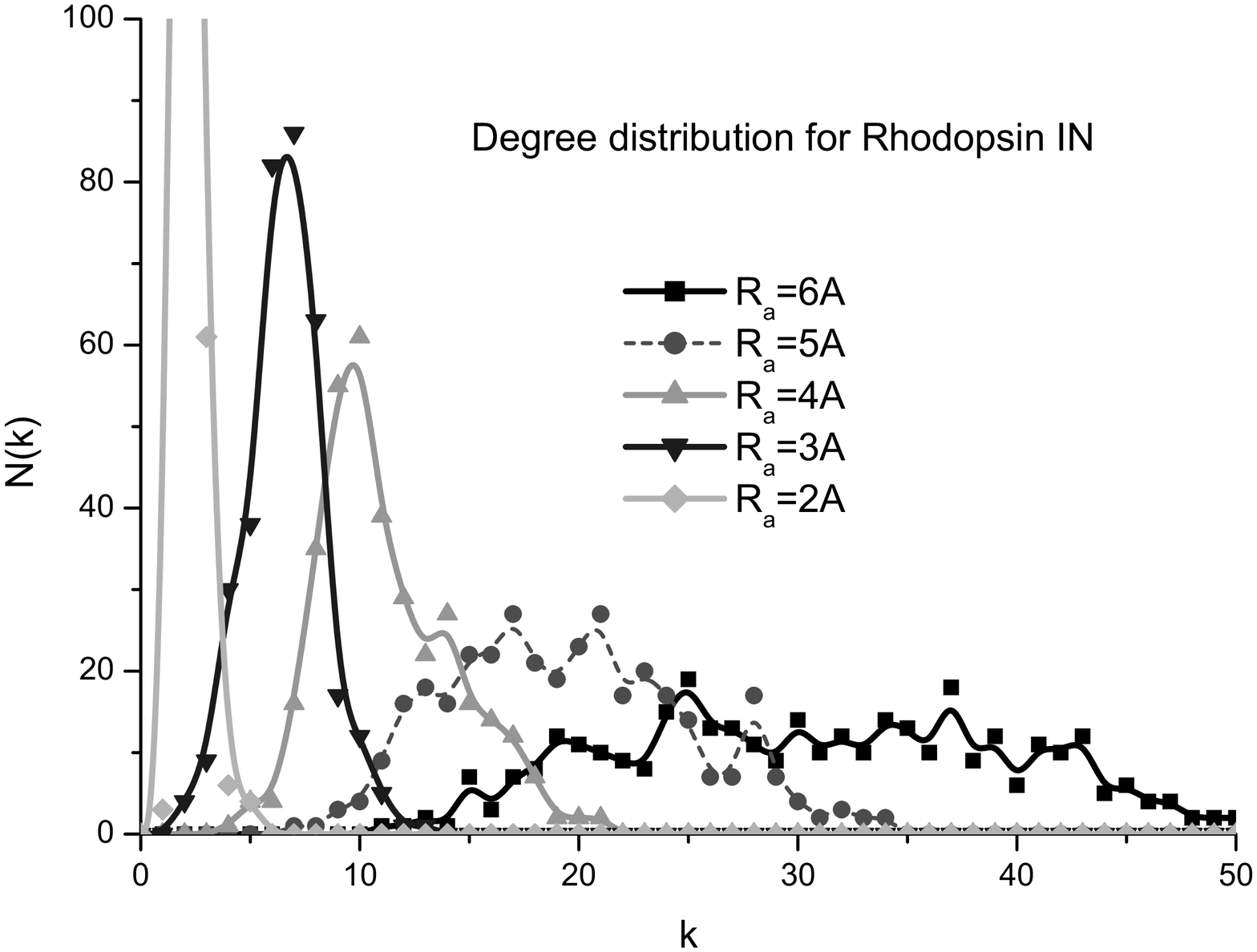}}
\clearpage
\resizebox{5.5in}{!}{\includegraphics{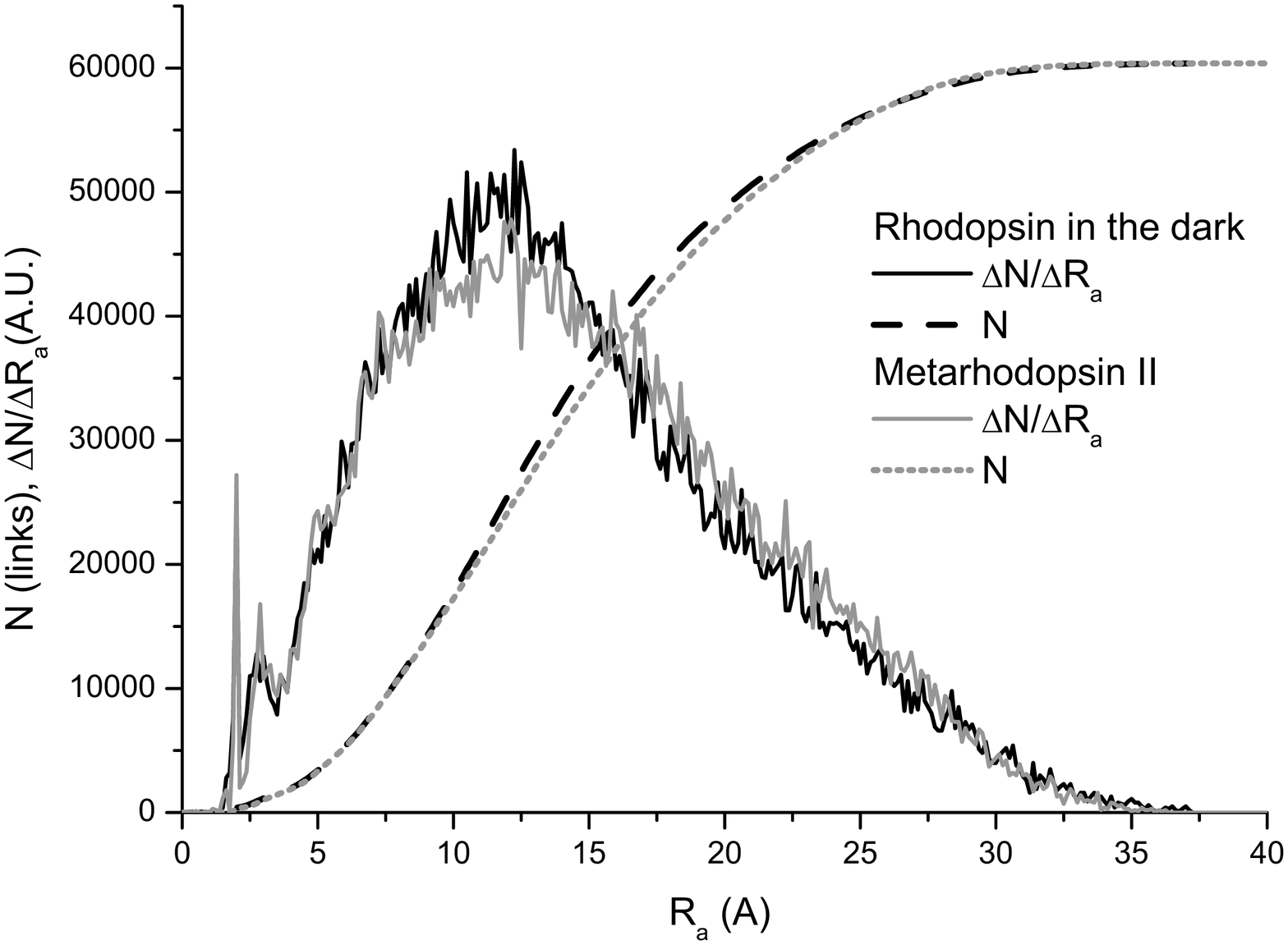}}
\clearpage
\resizebox{5.5in}{!}{\includegraphics{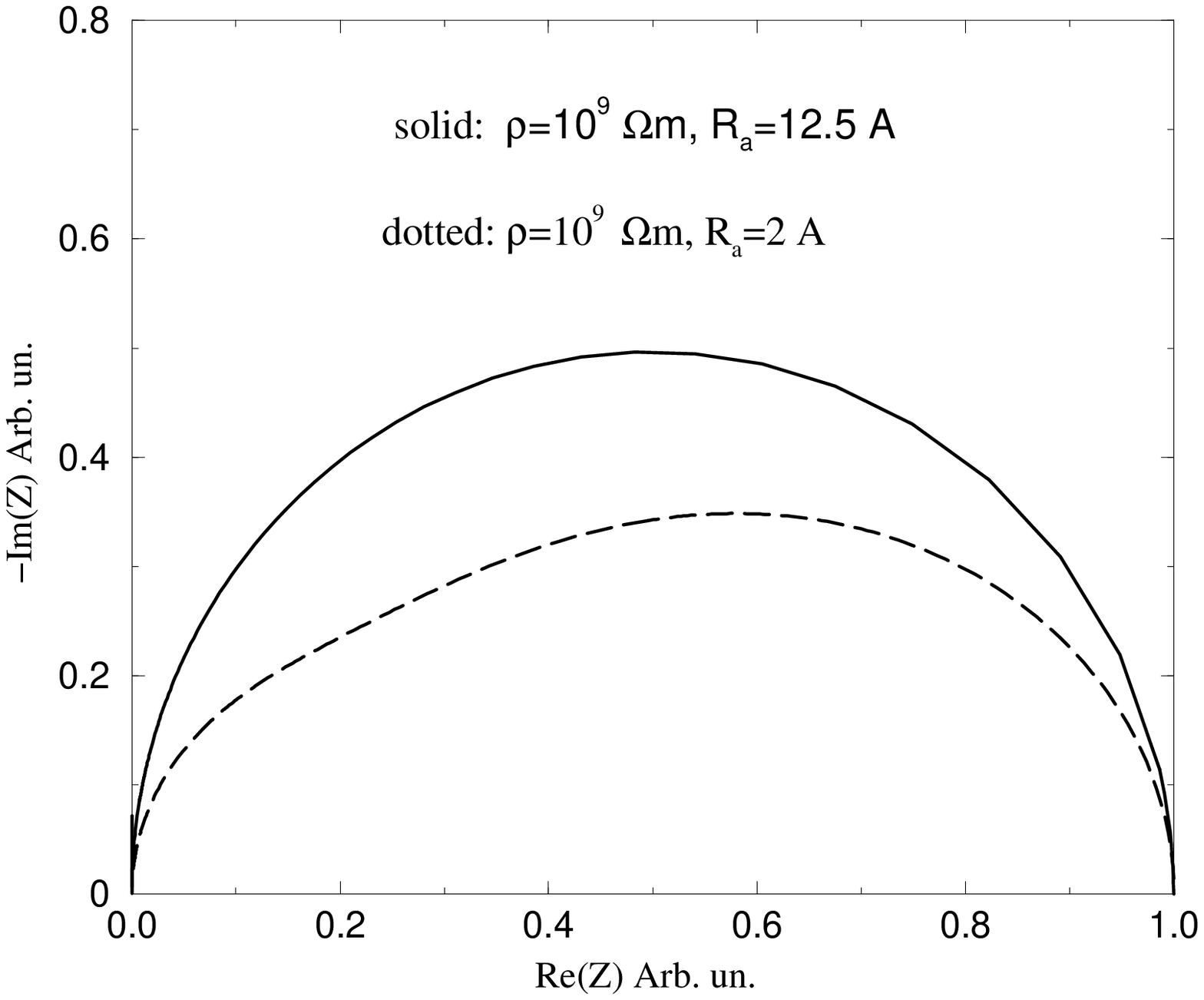}}
\clearpage
\resizebox{5.5in}{!}{\includegraphics{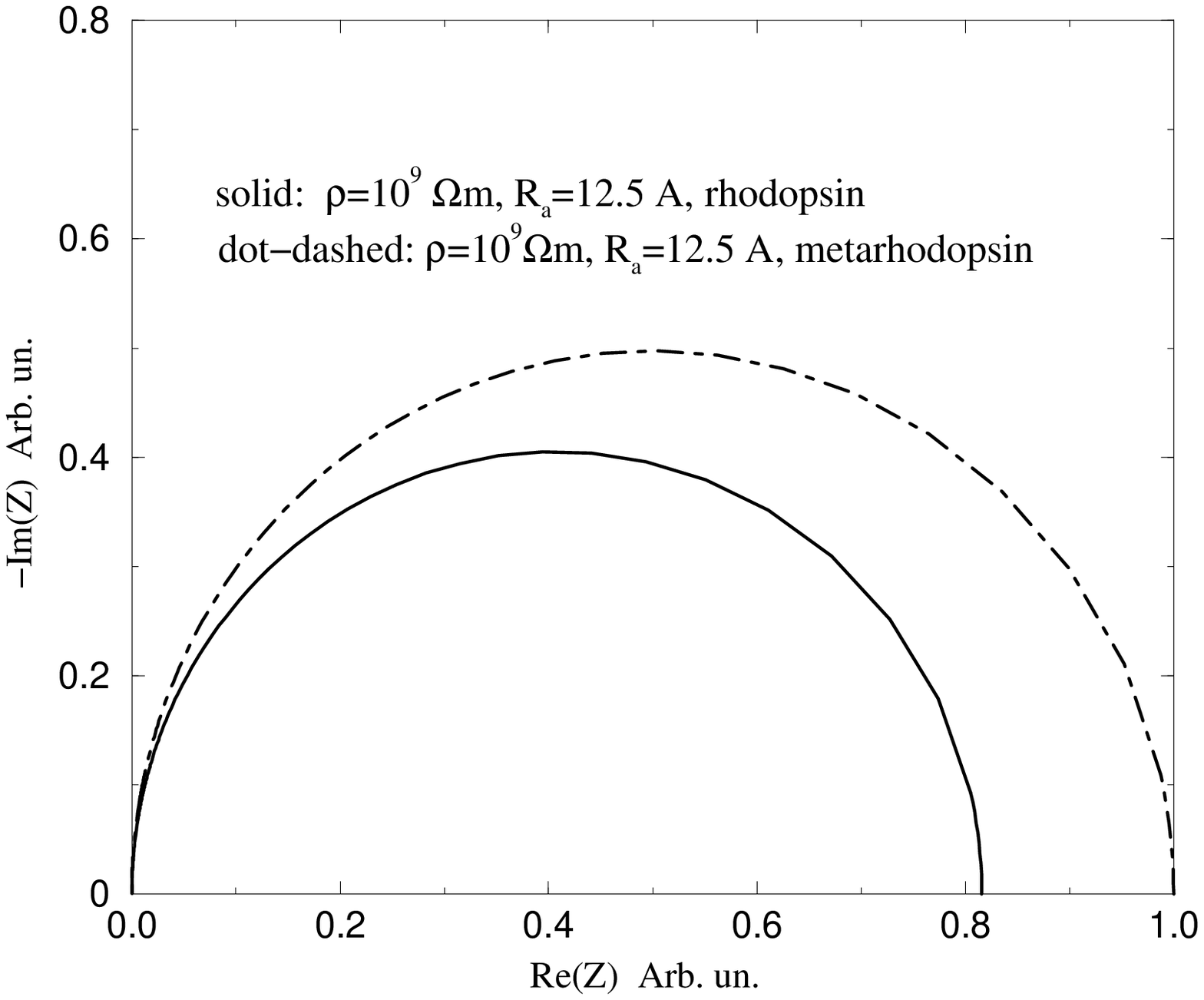}}
\clearpage
\resizebox{5.5in}{!}{\includegraphics{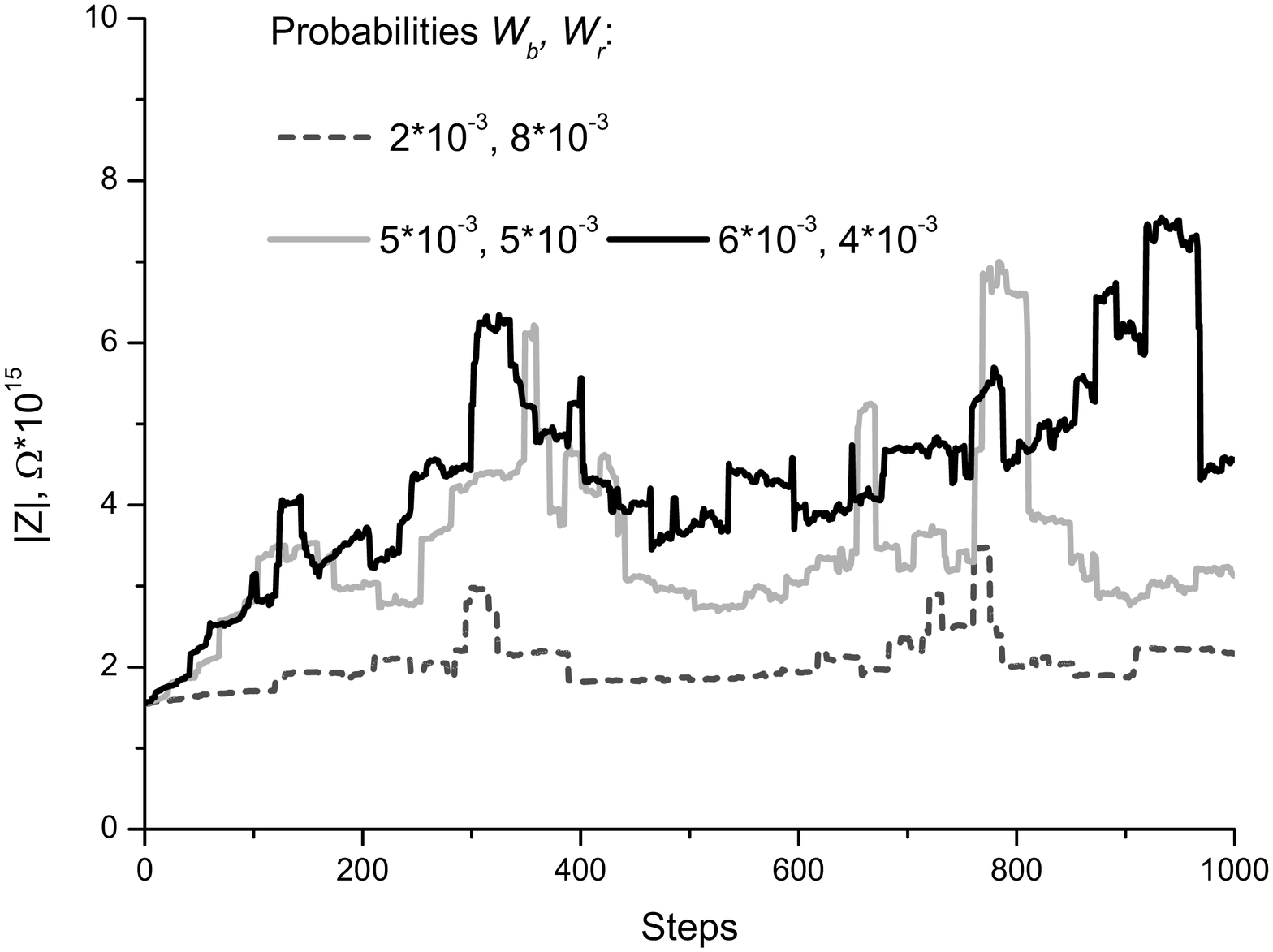}}


\begin{thebibliography}{99}
\bibitem{strogatz98} 
D.J. Watts and S.H. Strogatz, 
{\it Collective dynamics of 'small-world' networks}, Nature, 393,
		  440-442,1998

\bibitem{barabasi99}
A. L. Barabasi and R. Albert,
{\it Emergence of scaling in random networks}, 
Science,286,509-512,1999 

\bibitem{albert} 
R. Albert and A. L. Barabasi  , 
{\it Statistical mechanics of complex networks}, Rev. Mod. Phys.,74, 47-97,2002 
\bibitem{strogatz}
S.H. Strogatz, {\it Exploring complex networks}, Nature,410,268-276,2001 

\bibitem{jeong}
H. Jeong,B. Tombor,R. Albert,Z.N. Oltvai,A. L. Barabasi, {\it The large-scale organization of metabolic networks},Nature, 407, 651-654,2000  

\bibitem{callaway}
D.S. Callaway, M.E.J. Newman,S.H. Strogatz and D.J. Watts , 
{\it Networks robustness and fragility: percolation on random graphs}, 
Phys.Rev. Lett., 85,5468-71,2000 


\bibitem{goltsev}
A.V. Goltsev, S.N. Dorogovtsev and J.F.F. Mendes , 
{\it Critical phenomena in networks}, Phys. Rev. E, 67,026123-1-4, 2003

\bibitem{alon}
U. Alon, 
{\it Biological networks: the tinkerer as an engineer}, 
Science, 301,1866-1867, 2003 

\bibitem{argollo} 
M. Argollo de Menezes and A. L. Barabasi,
{\it Fluctuations in network dynamics}, Phys. Rev. Lett., 92, 028701-1-4,2004 

\bibitem{barabasi04} 
A. L. Barabasi and Z. N. Oltvai, 
{\it Network biology: understanding the cell's functional organization}, 
Nature Review,5, 101-113, 2004 

\bibitem{hladky} 
S.B. Hladky and D.A. Haydlon , 
{\it Discreteness of conductance change in biomolecular lipid  membranes in 
the presence of certain antibiotics}, 
Nature, 225, 451-453,1970


\bibitem{wu} 
Tzong-Zeng Wu , 
{\it A piezoelectric biosensor as an olfactory receptor for odour detection: 
electronic nose}, 
Biosensors and Bioelectronics,14, 9-18,1999


\bibitem{joachim}
C. Joachim and J.K. Gimzewski and A.Aviram , 
{\it Electronics using hybrid-molecular and mono-molecular devices}, 
Nature, 408, 541-548, 2000 

\bibitem{bayley}
H. Bayley and P.S. Cremer,
{\it Stochastic sensors inspired by biology}, 
Nature, 413,226-30,2001

\bibitem{firestein}
S. Firestein,
{\it How the olfactory system makes sense of scents},
Nature, 413,211-218,2001

\bibitem{xie}
Q. Xie,G. Archontis and S. S. Skourtis,
{\it  Protein electron transfer: a numerical study of tunneling through 
fluctuationg bridges}, 
Chem. Phys. Lett.,312,237-246,1999  

\bibitem{bezrukov}
S.M. Bezrukov and W. Winterhalter,
{\it Examining noise sources at the single-molecule level: 1/f noise of an 
open maltoporin channel}, 
Phys. Rev. Lett., 85, 202-205, 2000

\bibitem{lameh}
J. Lameh R. I. Cone, S. Maeda, M. Philip, M. Corbani, L. N\'adasdi, 
J. Ramachandran, G. M. Smith and W. Sad\'ee, 
{Structure and function of G protein coupled receptors}, 
Pharmaceutical Research, 7,1213-1221, 1990  

\bibitem{shacham}
S.Shacham,M.Topf,N. Avisar,F. Glaser,Y. Marantz,S. Bar-Haim,S. Noiman,Z. Naor and O. M. Becher,
{\it Modeling the 3D structure of GPCRs from sequence}, 
Medical Research Reviews,21, 472-483,2001 

\bibitem{vaidehi}
N. Vaidehi,W. B. Floriano,R. Trabanino,S. E. Hall,P. Freddolino,
		  E. J. Choi, G. Zamanakos and W. A. Goddard III, 
{\it Prediction of structure and function of G protein coupled receptors}, 
PNAS, 99, 12622-12627,2002  

\bibitem{elrod}
K. C. Chou and D. W. Elrod,
{\it Bioinformatical analysis of G-protein-coupled receptors},  
Journal of Proteome Research, 1,429-433, 2002  

\bibitem{spot}
SPOT - NOSED, Single Protein Nanobiosensor Grid Array, publisher V Framework Programme of European Community , 
{\it address http://www.nanobiolab.pcb.ub.es/projectes/Spotnosed/},2003-2005 

\bibitem{gether}
U. Gether and B.K. Kobilka,
{\it G protein-coupled receptors},
The Journ. Bio. Chem., 273,17979-17982,1998

\bibitem{pilpel}
Y. Pilpel and D. Lancet, 
{\it The variable and conserved interfaces of modeled olfactory receptor 
proteins}, 
Protein Science, 8, 969-977, 1999

\bibitem{crasto}
C. Crasto,M. S. Singer and G. M Shepherd,
{\it The olfactory receptor family album}, 
Genome Biology, 2,1027.1-1027.4,2001 

\bibitem{liu}
A. H. Liu,X. Zhang,G. A. Stolovitzky,A. Califano and S. J. Firestein, 
{\it Electronics using hybrid-molecular and mono-molecular devices},
Nature, 408,541-548,2000

\bibitem{palcz}
K. Palczewski,T. Hori,C. A. Behnke, H. Motoshima, B. A. Fox,
I. Le Trong, D. C. Teller, T. Okada,R. E. Stenkamp,M. Yamamoto and 
        M. Miyano,
{\it Crystal structure of Rhodopsin: A G protein-coupled receptor}, 
Science,289, 739-745, 2000 

\bibitem{yeagle}
P.L. Yeagle,G. Choi and A.D. Albert, 
{\it Studies on the structure of g-protein-coupled receptor rhodopsin
		  including the putative g-protein binding site in
		  unactivated and activated forms},
Biochemistry, 40,11932-11937,2001  

\bibitem{choi}
G. Choi,J. Landin, J. F. Galan, R.R. Birge, A.D. Albert and P.L. Yeagle , 
{\it Structural studies on metharhodopsin II, the activated form of the g-protein coupled receptor,rhodopsin},
Biochemistry,  41 ,7318-7324, 2002 
 
\bibitem{sakmar}
T.P. Sakmar, S.T. Menon, E.P. Marin and E.S. Awad,
{\it RHODOPSIN: Insights from recent structural studies}, 
Annu. Biomol. Struct.,31 , 443-482,2002  

\bibitem{pdb}
  Research Collaboratory for Structural Bioinformatics ,
    Protein data bank ,
publisher State University of New Jersey , 
{\tt address http://www.rcsb.org/pdb} 

\bibitem{yang}
H. Yang,G. Luo,P.Karnchanaphanurach,T.M. Louie,I. Rech,S. Cova,L. Xun and X. S. Xie, 
{\it Protein conformational dynamics probed by single-molecule 
electron transfer}, Sciences, 302, 262-266,2003

\bibitem{song}
Xueyu Song,
{\it An inhomogeneous model of protein dielectric properties:
		  intrinsic polarizabilities of amino acids},
J. Chem. Phys.,116 ,9359-9383,2002  


\bibitem{kobilka2}
B. Kobilka,U. Gether,M. Seifert,S. Lin and P. Ghanouni, 
{\it Examination of ligand-induced conformational changes in the beta
		  2 andrenergic receptor},
Life Sciences,62,1509-1512, 1998

\bibitem{kobilka}
B. Kobilka,U. Gether,M. Seifert,S. Lin and P. Ghanouni, 
{\it Characterization of ligand-induced conformational states in the beta 
2 andrenergic receptor}, 
J. Receptor and Signal Transduction Reseach, 19, 293-300,1999

\bibitem{rammal}
R. Rammal,C. Tannous and A.M.S. Tremblay, 
{\it 1/f noise in random resistor networks: 
fractals and percolating systems}, 
Phys. Rev.A,31,2662-71,1985  


\bibitem{prl85}
C. Pennetta, G. Trefan and L. Reggiani,
{\it Scaling and universality in stationary random resistor networks},
Phys. Rev.Lett., 85,5238-5241,2000            

\bibitem{stauffer}
D. Stauffer and A. Aharony,
{\it Introduction to Percolation Theory},
publisher  Taylor and Francis,address London, 1992 

\end{thebibliography}
\end{document}